\documentclass[%
 reprint,
 amsmath,amssymb,
 aps,
 superscriptaddress,
]{revtex4-2}

\usepackage{graphicx}
\usepackage{dcolumn}
\usepackage{bm}
\usepackage{hyperref}

\usepackage[dvipsnames]{xcolor}

\definecolor{myblue}{RGB}{0, 114, 189}
\definecolor{mymaroon}{RGB}{217, 83, 25}
\definecolor{myyellow}{RGB}{237, 177, 32}

\begin{document}

\preprint{APS/123-QED}

\title{Testing Quantum Dissipation Theory with Electron Diffraction}

\author{Raul Puente}
\affiliation{%
 Facultad de Ciencias, Universidad Nacional Aut\'{o}noma de M\'{e}xico, M\'{e}xico}%
\author{Zilin Chen}%
\affiliation{%
 Department of Physics and Astronomy, NorthWestern University, Evanston}%
\author{Herman Batelaan}
\affiliation{%
 Department of Physics and Astronomy, University of Nebraska, Lincoln}%


\date{\today}

\begin{abstract}
Decoherence can be provided by a dissipative environment as described by the Caldeira-Leggett equation. This equation is foundational to the theory of quantum dissipation. However, no experimental test has been performed that measures for one physical system both the dissipation and the decoherence. Anglin and Zurek predicted that a resistive surface could provide such a dissipative environment for a free electron wave passing close to it. We propose that the electron wave's coherence and energy loss can be measured simultaneously by using Kapitza-Dirac scattering for varying light intensity.
\end{abstract}

\maketitle


\section{Introduction}

\begin{figure}[ht]
   \includegraphics[width = \linewidth]{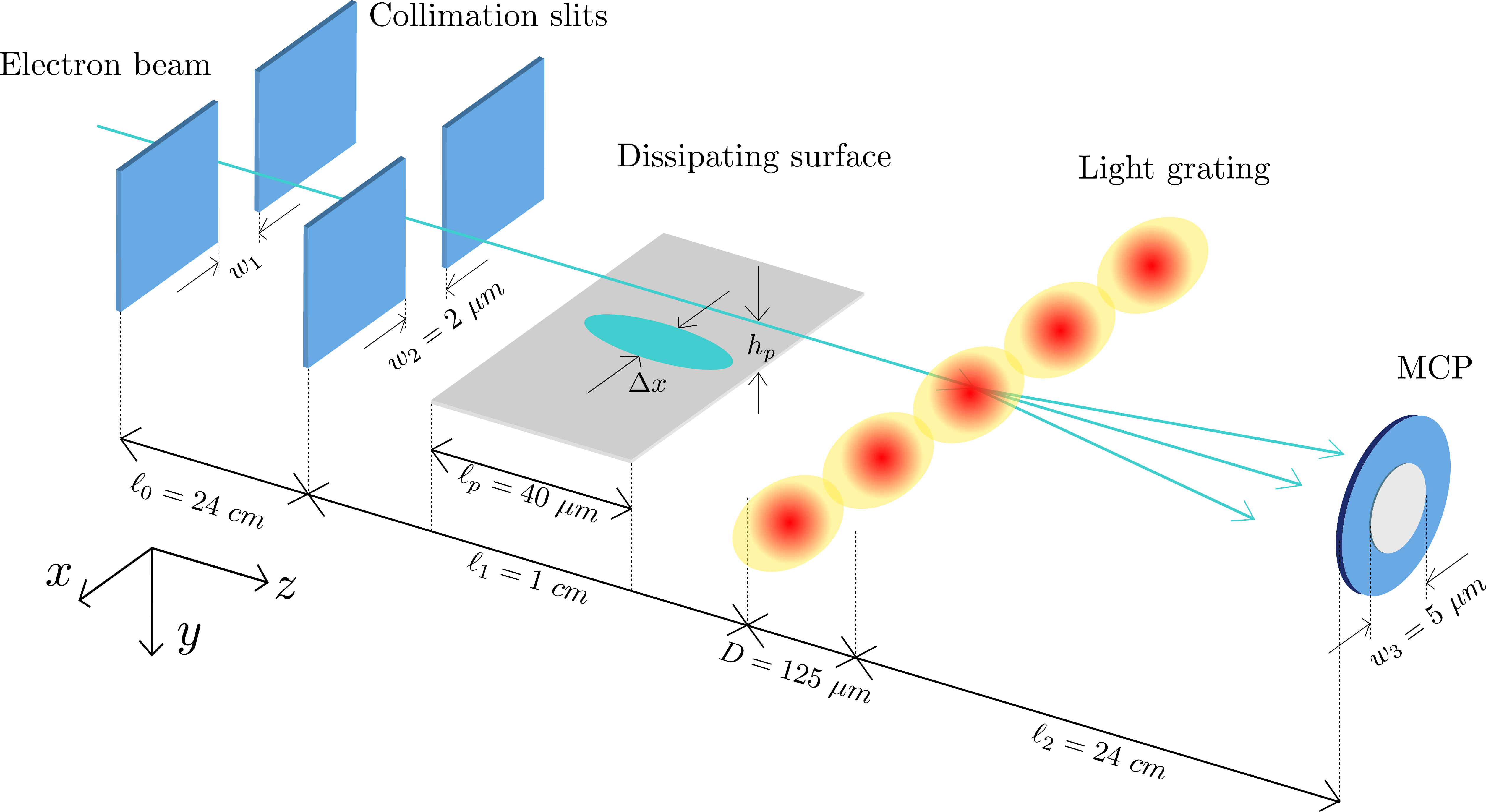}
   \caption{Thought experiment. (a) The electron beam (blue line) starts at the left top and travels to the bottom right. Two collimating slits define the electron beam and its transverse coherence length, $\Delta x$. The electron beam passes above and parallel to the surface at a height $h_{P}$. The electrons that have lost coherence and energy through their interaction with the surface, intersect a standing wave of light (yellow) and are diffracted by the Kapitza-Dirac effect. The image of the diffracted beam is observed in the far-field.}
  \label{fig:setup}
\end{figure}

The process of quantum dissipation shares with quantum decoherence its purpose to find a smooth transition from quantum mechanics to classical mechanics \cite{Zeh1970,Leggett1987,Zurek1991}. To what extent quantum dissipation and decoherence address the quantum measurement problem is an active research topic \cite{schlosshauer,Zurek2014}. In quantum dissipation, physical systems are studied quantum mechanically whose classical counterpart dissipate energy. The prototypical system is Brownian motion; a larger particle interacts with a large bath of surrounding and interacting smaller particles. The general prediction by Caldeira and Leggett is that the larger particle not only looses kinetic energy, but also decoheres. Moreover, the rates of both are linked together in a predictable fashion. The appeal of the Caldeira-Leggett theory \cite{Cald1983} stems partly from its ubiquitous nature and could thus, in principle, be applied broadly. 

Arndt et al. \cite{hackermuller2004,arndt2005} observed decoherence of interfering molecules caused by a thermal heat bath. The molecular internal states are closely spaced so that they can be excited by the black-body radiation of the environment when it is heated. Hornberger solved the Hund's paradox by considering the relative rate of collisions versus tunneling of molecular enantiomeres. The collisions decohere the molecular enantiomeres before a superposition of right and left handed molecules can evolve by tunneling \cite{trost2009}. Both of these beautiful textbook experiments exhibit decoherence due to a environmental bath. However, even if many collisions cause decoherence as in the Brownian-motion models, dissipation is not considered. 

The Caldeira-Leggett has not been verified experimentally yet in the sense of a simultaneous measurement of the decoherence rate and dissipation rate, so that their linkage could be verified. Zurek proposed the study of electrons flying above a conductive surface \cite{Anglin97} as a decohering environment. As the electron passes close to the surface it sets up an image charge. The image charge is comprised of a large number of redistributed electrons, which provide a noisy environment. The retardation of the field of the induced image charge, makes it lag behind the position of the passing electron and is thought to cause dissipation \cite{boyer74}. The rate of dissipation is used to predict the decoherence rate \cite{Anglin97}. The initial reported observation of the decoherence effect \cite{Sonnentag07} was criticized in \cite{Beierle18,Kerker2020}. If the electron-wall system is a good candidate for observing decoherence is a topic of current research \cite{Kerker2020,zilin2020}. Regardless of the outcome of that research, no simultaneous observation of energy dissipation and decoherence has been reported. Given that at a clear predicted decoherence signal, the energy loss is small and hard to observe this is not surprising.  

Here we propose the use of the Kapitza-Dirac effect as an analysis tool for the electron wave emerging after interaction with a conductive wall. The Kapitza-Dirac effect is the diffraction of electron from a standing wave of light \cite{batelaan2007} and has been observed in the Raman-Nath \cite{freimund2001} and Bragg \cite{freimund2002} regime. The central idea is that the electron energy changes the position of the diffracted peaks and that this change in position scales with the order of the diffraction peak. Thus the light-grating can be used to measure the electron's energy loss sensitively. The electron's transverse coherence length determines the contrast of the diffraction pattern. The decoherence process reduces the transverse coherence length, and thus the contrast of the observed diffraction pattern. Thus, the observation of the diffraction peak position and the diffraction contrast constitute a simultaneous measurement of decoherence and dissipation.   

We now proceed to work out the basic theory to find the experimental parameters needed to do such a test and conclude that these parameters are within reach of current capabilities.    

\section{Theory}

\subsection{Decoherence and energy loss}
Anglin and Zurek's model~\cite{Anglin97} give the decoherence time for an electron interacting with a resistive semiconductor wall, 
 \begin{equation}
 \tau_{\mathrm{dec}}^{\text {Zurek }}=\frac{4 h^{2}}{\pi e^{2} k_{\mathrm{B}} T \rho} \frac{z^{3}}{(\Delta x)^{2}},
 \label{eqn:Zurek}
 \end{equation}
where $h$ is Planck's constant, $k_{\mathrm{B}}$ is Boltzmann's constant, $T=300$ $\mathrm{K}$ is the temperature, $\rho$ is the resistivity of the wall, $z=h_{P}$ is the electron height above the surface, $\Delta x$ is the transverse coherence length or in other words, the distance between two interfering paths. For the proposed method, a correction factor is needed. The image charge associated with one electron path can overlap with the image charge distribution associated with another electron path. The overlap removes ``which-way'' information by a factor $C=(z/\Delta x)^{2}$~\cite{ZurekPC}. The decoherence amount is given by $R_{dec}=\int{dt(C\tau_{dec}^{Zurek})^{-1}}$. The dissipation energy loss is given by~\cite{boyer74}:
\begin{equation}\label{elossrate}
P=\frac{e^{2} \rho v^{2}}{16 \pi z^{3}}.
\end{equation}
The energy loss $\Delta E$ is $P\times t_{f}$ (for $\Delta E<<E$), where $t_{f}$ is the flight time of the electron over the surface, and the kinetic energy is $E=mv^2/2$. For completeness, we note that the decoherence amount
\begin{equation}
R_{dec}=\left(\frac{\Delta x}{\lambda_{\mathrm{th}}}\right)^{2} \frac{\Delta E}{mv^{2}},
\end{equation}
has a fixed relation to the energy loss and can thus be measured.

\subsection{Kapitza-Dirac effect}
In a reversal of the traditional role of a light wave diffracting from a material structure, Kapitza and Dirac's \cite{kapitsa33} proposal to diffract electrons from a standing wave of light was realized in 2001  \cite{freimund2001}. To find the interaction potential it suffices to study the classical motion of an electron in a standing electromagnetic wave. Consider an electron with velocity $\vec{v}=v\hat{z}$ and a standing wave vector potential 
\begin{equation}
    \vec{A}=\frac{E_0}{2\omega}\cos(kx)\sin(\omega t)\hat{z}.
\end{equation}
The fields are given by
\begin{equation}
    \vec{E}=E_0\cos{kx}\cos{\omega t}\hat{z},
\end{equation}
and magnetic component
\begin{equation}
    \vec{B}=-\frac{1}{c}E_0\sin{kx}\sin{\omega t}\hat{y}.
\end{equation}
 The phase differences between the fields rectify the oscillatory forces so that the electron experiences a time averaged ponderomotive potential
 \begin{equation}
    V_p=\frac{q^2I}{2m\epsilon_0c\omega^2}\cos^2{kx}=V_0\cos^2{kx},\label{ponderomotive}
\end{equation}
where $I=\epsilon_0cE^2_0/2$ is the intensity of the laser. This potential writes a periodic phase pattern on the electron wave which is used in the electron diffraction calculation.  

\subsection{Single--electron diffraction}

The diffraction pattern due to the Kapitza-Dirac effect can be obtained using Feynman's path integral approach \cite{feynman48}. This method provides convenient control over the amount of coherence present in the electron's quantum mechanical motional state.

We start with the initial wave function described by a source point,
\begin{equation}\label{incoherent}
    \Psi_{\text{inc}}(x''',\chi)=\delta(x'''-\chi),
\end{equation}
where $x'''$ represents the position at the source plane, $\chi$ represents the source point. Next we propagate from the source plane to the second slit using the kernel
\begin{equation}
    K_{x_a\to x_b}(x_a,x_b)=e^{i\frac{2\pi}{\lambda_{\text{dB}}}\sqrt{(x_a-x_b)^2+\ell_{x_a\to x_b}^2}}.
\end{equation}
The wave function becomes
\begin{equation}
    \Psi_{\text{slit}}(x'',\chi)=\int_{-\infty}^{\infty}{K_{x'''\to x''}(x''',x'')\Psi_{\text{inc}}(x''',\chi)\ dx'''},
\end{equation}
where $x''$ represents a position at the slit plane, and the kernel is given by
with $\ell_{x'''\to x''}$ being the distance between the source point and the location at the second slit. The second collimating slit has a Gaussian transmission profile to avoid edge effects. The propagation from the second slit plane to right before the laser region is given by
\begin{equation}
    \Psi_{bl}(x',\chi)=\int_{-\infty}^{\infty}{K_{x''\to x'}(x'',x')\Psi_{\text{slit}}(x'',\chi)\ dx''}.
\end{equation}
While crossing the laser region, there electron wave after interacting with the laser is
\begin{equation}
    \Psi_{al}(x',\chi)=\Psi_{bl}(x',\chi)e^{i\phi_l(x')},
\end{equation}
where the phase is
\begin{equation}
    \phi_l(x')=-\frac{qV_p(x')t_\ell}{\hbar},
\end{equation}
where $q$ the charge of the electron, $V_p$ is given by \eqref{ponderomotive}, and $t_\ell=D/v_e$ is the time it takes the electron with velocity $v_e$ to cross the laser region of width $D$. Finally, the electron wave is propagated to the detection screen plane,
\begin{equation}
    \Psi_{\text{detect}}(x,\chi)=\int_{-\infty}^\infty{K_{x'\to x}(x',x)\Psi_{\text{slit}}(x',\chi)\ dx'}.
\end{equation}
With $x$ representing the position at the screen. The probability on the detection screen is given by
\begin{equation}
    P_{\text{detect}}(x,\chi)=|\Psi_{\text{detect}}(x,\chi)|^2,
\end{equation}
which gives the diffraction pattern produced by a single source point at $\chi$ (eq. \eqref{incoherent}). 

To control the amount of coherence, the diffraction pattern probability distribution is integrated over the source point distribution. This incoherent sum reduces the contrast of the diffraction pattern. Because the incoherent sum can be performed at any intermediate plane, the source width controls the amount of coherence at any location. The source shape is chosen to be a Gaussian profile with standard deviation $\sigma_e$. A finite detection size is modeled by convolution with a detection slit of width $\sigma_{d}$,
\begin{equation}
    P_{\text{final}}(x)=e^{-\frac{x^2}{2\sigma_d^2}}\ast\int_{-\infty}^\infty{e^{-\frac{\chi^2}{2\sigma_e^2}}P_{\text{detect}}(x,\chi)\ d\chi}.
\end{equation}

\begin{figure*}[t]
  \centering
  \includegraphics[width=0.5985\textwidth]{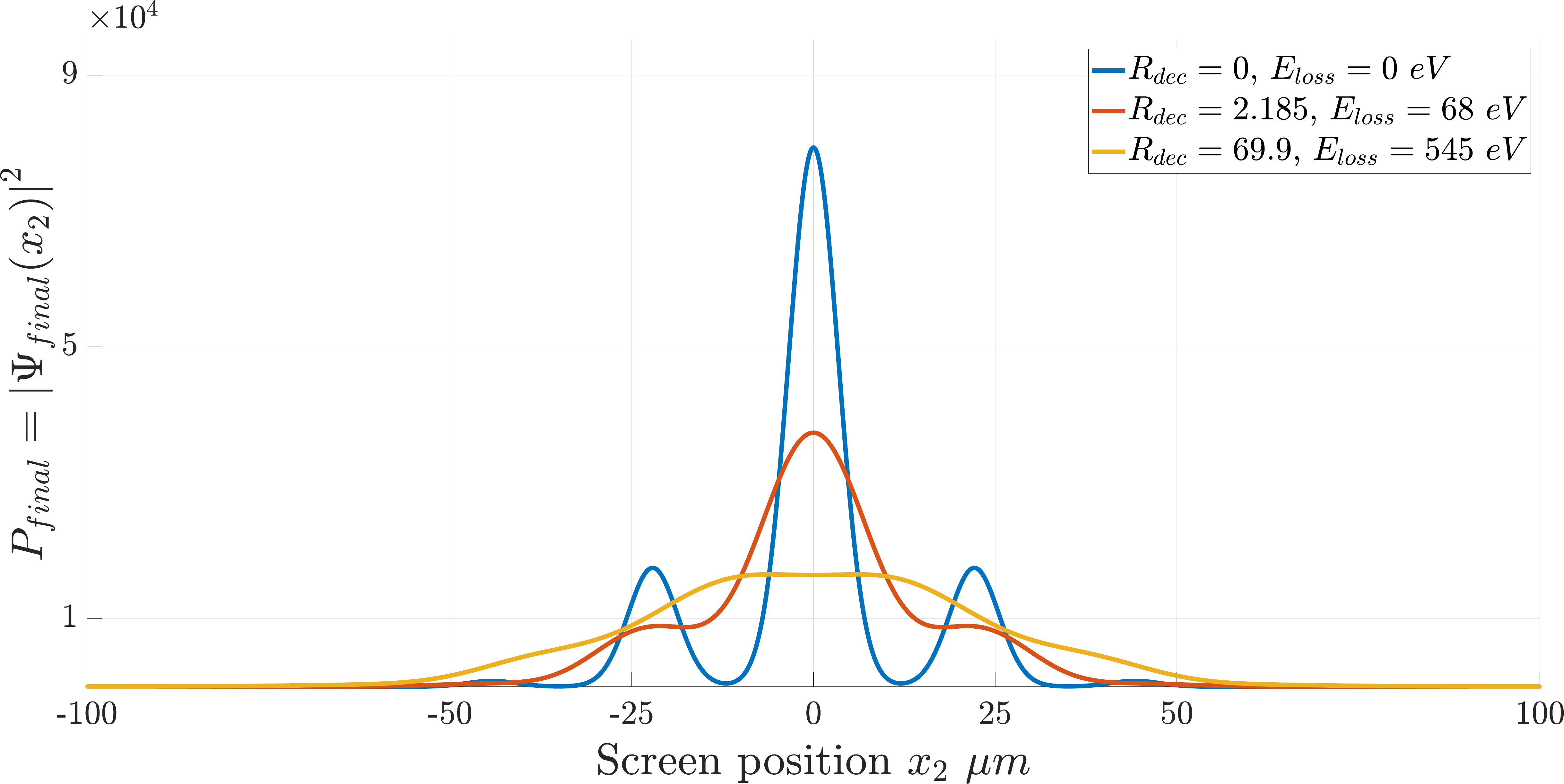}
  \caption{Low laser intensity. Diffraction patterns are obtained for the K-D effect at a laser intensity of $I=1\times10^{14}\ W/m^2$ and a varying transverse coherence length and energy loss. The coherence length of the electron beam was reduced by increasing the incoherent source width $w_1$ to simulate an increasing decoherence amount while the energy of the electron beam was lowered to simulate energy loss. In an experiment, both are predicted to occur when the height $z=h_p$ of the electron above the conductive plate is reduced. The line in blue represents the pattern for no plate (i.e. no decoherence and no energy loss). The line in maroon represents the pattern for a height $h_p=2\ \mu m$ to give a decoherence amount of 2.185 and an energy loss of $68\ eV$. The line in yellow represents the pattern for a height $h_p=1\ \mu m$ to give a decoherence amount of 70 and an energy loss of $545\ eV$ (color figure online). As the decoherence amount grows, the diffraction pattern contrast diminishes. The energy loss is not obviously recognizable.}
  \label{low_intensity}
\end{figure*}
\begin{figure*}[t]
  \centering
  \includegraphics[width=0.5985\textwidth]{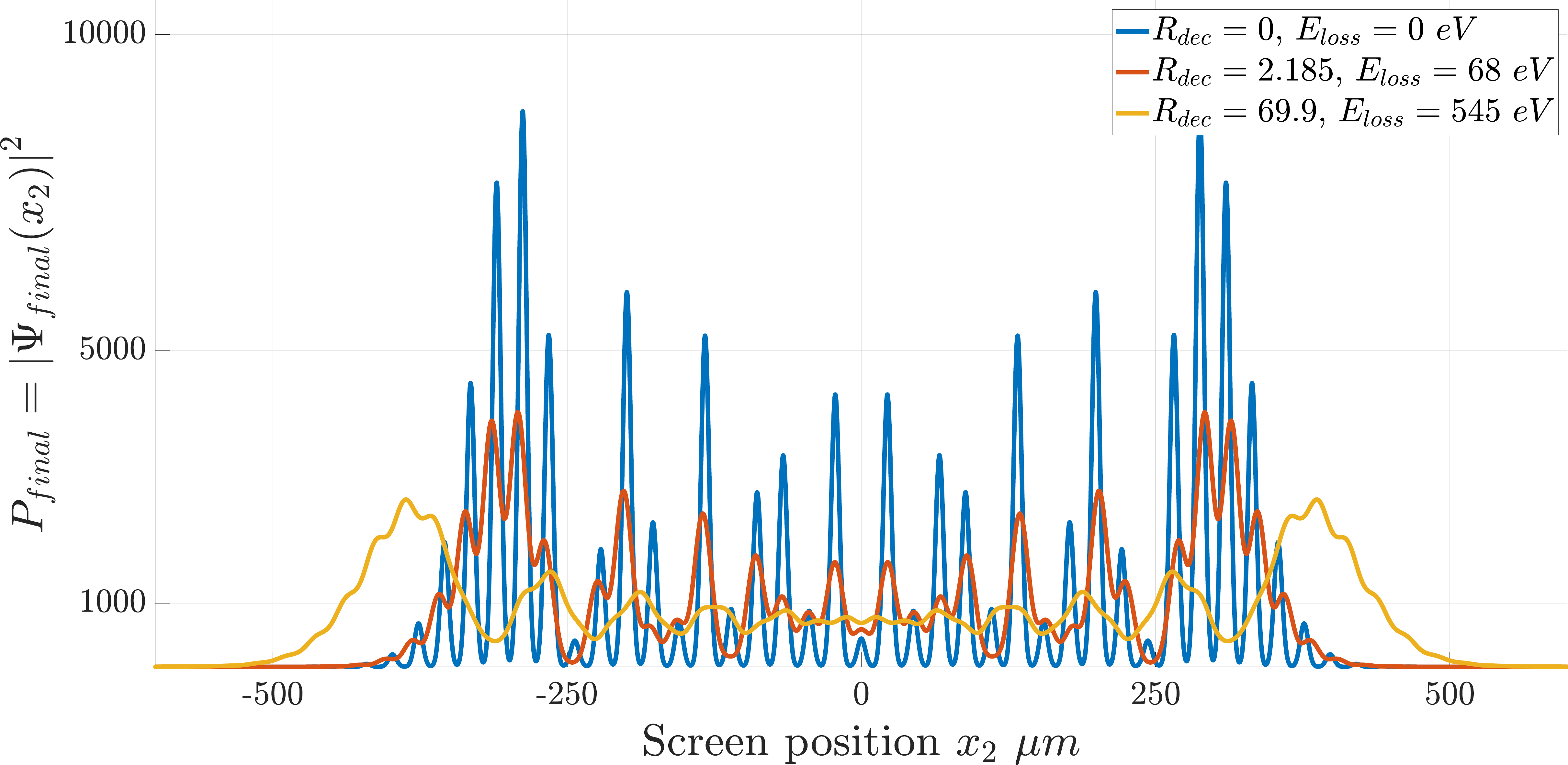}
  \caption{High laser intensity. Diffraction patterns are obtained for the K-D effect at a laser intensity of $I=18\times10^{14}\ W/m^2$ and a varying decoherence amount and energy loss. The variation and the parameter values are identical to that for Fig. \ref{low_intensity}. As the decoherence amount $R_{dec}$ grows, the diffraction pattern contrast diminishes. As the dissipation amount grows, the diffraction peaks shift outward; the yellow maxima occur at a larger absolute screen position values than their blue counterpart.}
  \label{high_intensity}
\end{figure*}

The FWHM of the anti-diagonal of the density matrix is used as the measure of the transverse coherence length. In the position representation, the density matrix for one source point is given by $\rho_{P}= \langle x_{i}|\Psi\rangle\langle\Psi|x_{j}\rangle$. At the location right before the laser, the incoherent sum is given by

\begin{equation}
\rho_{in}=\int{\langle x_{i}|\Psi_{bl}\rangle\langle\Psi_{bl}|x_{j}\rangle\mathrm{~d} \chi}.
\end{equation}

As the source width is increased, the FWHM of the anti-diagonal decreases, consistent with Heisenberg uncertainty principle. The initial source width (that is the first slit width) sets the coherence length of the electron's density matrix $\rho_{in}(x_{i},x_{j})$ just before it enters the standing wave of light. This density matrix is affected by the source width to model the effect of decoherence as described by the ``decohered'' density matrix $\rho_{dec}(x_{i},x_{j})$.
The ratio of the coherence terms in these density matrices at a location $(x_{i},x_{j})=(x+\frac{1}{4}\sqrt{2}\Delta x,x+\frac{1}{4}\sqrt{2}\Delta x)$ is given by 

\begin{equation}
\frac{\rho_{dec}(x_{i},x_{j})}{\rho_{in}(x_{i},x_{j})}=\exp\left(-\frac{t_{flight}}{C\tau_{Zurek}(\Delta x)}\right).
\end{equation}
The transverse coherence length, $x_{coh}$, is found by solving
\begin{equation}
\exp\left(-\frac{t_{flight}}{C\tau_{Zurek}(x_{coh})}\right)=1/2.
\end{equation}

\begin{table*}[htbp]
\caption{\label{tab:decoherence rate} The decoherence amount and energy loss are given for three models.}
\begin{ruledtabular}
\begin{tabular}{ccccccc}
\multicolumn{7}{c}{Illuminated gallium arsenide}\\ \hline
Figure number & \multicolumn{3}{c}{2} & \multicolumn{3}{c}{3}\\
Figure color & blue & maroon & yellow & blue & maroon & yellow\\
\hline
Laser intensity $I\ W/m^2$  & $1\times10^{14}$ &  $1\times10^{14}$& $1\times10^{14}$ & $18\times10^{14}$ & $18\times10^{14}$ & $18\times10^{14}$\\
Height $h_p\ \mu m$ & $\infty$ &$2$ & $1$ & $\infty$ & $2$ & $1$\\
Slit width $w_1\ \mu m$ & $6.7$ & $22$ & $100$ & $6.7$& $22$ & $100$ \\
\hline\\
\multicolumn{7}{c}{Zurek's model}\\ \hline
Energy loss $\Delta E\ eV$& 0 & 68 & 545 & 0 & 68 & 545 \\
Decoherence amount\footnote{The ratio of time of flight over decoherence time. All values are calculated with correction C.} $R_{dec}$
& 0 & 2.185 & 69.9 & 0 & 2.185 & 69.9 \\ 
\end{tabular}
\end{ruledtabular}
\end{table*}


\section{Results, Discussion and Conclusion}
\subsection{Results}
The parameters are chosen within the range achieved in previous experiments (Fig. \ref{fig:setup}). The $2.5\ keV$ electron beam passes through a first slit of width $w_1$ (varied from 6.7, 22 to 100 microns for the results in Figs. \ref{low_intensity} and \ref{high_intensity}) and a second collimating slit of 1 micron placed 24 $cm$ downstream. A gallium arsenide plate of 40 microns of length placed 1 $mm$ after the second slit is illuminated with $<800\ nm$ light is located underneath of the diffracted electron beam at a variable height $h_p$. The intensity  of the illumination is used to excite the band gap, populate the conduction band and thus decrease its resistivity to a value of $144\ \Omega m$. Note that not-illuminated GaAs has a resistivity exceeding $10^6\ \Omega m$ (vertical-gradient-freeze grown undoped GaAs crystals with 110-crystalline orientation is commercially available with 2.5-5.1 $10^6\ \Omega m$) and several $mW$ of light can easily reduce the resistivity to $5\ \Omega m$ \cite{zilin_arxiv}. This provides a convenient continuous control of the resistivity. The time of flight is $t_f=1.3\times10^{-12}\ s$. Then, after a downstream distance of $1\ cm$ the electron beam interacts with a standing wave of light from a $532\ nm$ wavelength laser with a laser beam waist of 125 microns placed perpendicular to the initial direction of the beam. After a further distance of $24\ cm$ a 5 micron detection slit was used in the simulation.

A diffraction pattern was obtained for a laser intensity of $I=1\times10^{14}\ W/m^2$, which is shown in  the curve in blue in Fig. \ref{low_intensity}. The decoherence amount was varied by changing the incoherent source width $w_1$ (see above) to simulate a loss of coherence due to the interaction of the electron beam with the metal plate placed 1 or 2 microns underneath the beam (Fig. \ref{low_intensity}). The time it takes the electron to cross the laser region is $t_\ell=4.2\times 10^{-12}\ s$ for $h_p=2\ \mu m$ and $t_\ell=4.8\times 10^{-12}\ s$ for $h_p=1\ \mu m$.
The decoherence amount was measured using the FWHM of the gaussian distribution of the density matrix anti-diagonal. For the incoherent source with $w_1$ of $6.9\ \mu m$, the FWHM occurred at $\pm200\ \mu m$. After changing the width $w_1$ to $22\ \mu m$, the value of the gaussian at $\pm200 \mu m$ was approximately $\exp{(-2.185)}$ smaller, i.e. there was a decoherence amount of 2.185. The energy of the beam was set to lower value according to  eq. \eqref{elossrate} in order to take into account both the decoherence and the energy loss after the interaction of the beam with the metallic plate. This is shown in the same figure with the curve in maroon. This curve shows how the lateral peaks of the diffraction pattern are barely visible due to decoherence. This was done again, but now with a decoherence amount of 70 caused by the plate now being placed at a height $h_p$ of 1 micron underneath the beam, which is shown in the same figure yellow. This curve now shows how the diffraction pattern has been completely lost.\\[0.1cm]
Fig. \ref{high_intensity} shows the same procedure described before, but now with a laser intensity of $I=18\ W/m^2$. This was done in order to be able to see more diffracting orders, as the energy loss is more visible as the diffraction order increases. Indeed, the 13th peak of the diffraction pattern with a 2.185 decoherence amount shifts by less than 2\%, while the pattern obtained with a decoherence amount of 69.9, there is a shift of almost 35\%.

\subsection{Discussion and Conclusion}
Not only can the dissipation rate and decoherence rate be measured in one physical system, but one measurement (Fig. \ref{high_intensity}) suffices to test the Caldeira-Leggett prediction. This is experimentally advantageous, as the effect of rate changes or other experimental drifts are ameliorated. Assuming that the proposed parameters regime can be reached and that dissipation and decoherence play their predicted role, the physical system can be varied in interesting ways. If the surface chosen is a semi-conductor then electrons in the valence band can be excited to the conduction band to lower the resistivity and thus develop decoherence-free surfaces to act as a wave guide for electrons \cite{zilin_arxiv}. For thin surfaces the excitation can proceed through back-illumination \cite{Sturge62}.

An alternative way to the use of the Kapitza-Dirac effect is the use of nano-fabricated gratings. They have been show to support the detection up to the $20^{th}$ diffraction order \cite{Barwick06}. However, the detection rate at higher orders was low for the typical $0.5$ open fraction gratings ($50\ nm$ slit, $100\ nm$ periodicity). To overcome that problem, nano-fabricated grating with narrower slits would be needed. The single slit diffraction pattern acts as the envelope for the grating diffraction pattern. For narrower slits the envelop broadens and enhances the higher diffraction orders. It appears that $10\ nm$ wide slit is within current nano-fabrication capabilities, and our proposed method, encourages such work to be performed.

The choice of placing the surface before or after the light grating can also be considered. It may appear natural to place the surface after the light grating when considering two paths starting at the grating that interfere in the detection plane. However as each point source at the collimating slit illuminates the grating coherently, which in turn recombines the electron paths, the thought experiment is an interferometer with separate paths. At the location where the paths are separated the most is where the surface causes most decoherence, regardless if that is before are after the grating. 

The proposed method could fail to test the predicted relation between the dissipation and the decoherence rate, if the classical dissipation rate estimate is too high.  This could be compensated by a closer proximity to the wall. On the other hand the length scale could become so small that for all practical purposes the quantum regime is reached and an estimate of a quantum dissipation rate is needed \cite{Scheel12,Machnikowski06}.  

Finally, we note that it may be of interest to theoretically investigate the decoherence of ion beams by nearby conducting surface to establish if that will also make the testing of quantum dissipation accessible. 
\vspace{0.6cm}
\begin{acknowledgments}
 We gratefully acknowledge support by the U.S. National Science Foundation under Grant No. PHY-1912504.
\end{acknowledgments}


\bibliographystyle{apsrev4-2}
\bibliography{apssamp}

\end{document}